\documentclass[a4paper,twosides]{article}
\usepackage{CJK,multicol,multirow,graphics,fancyhdr,epstopdf}
\usepackage{amsmath,amsfonts,amssymb,bm,upgreek,mathrsfs,ccmap,mathcomp}
\usepackage[pagewise,switch,columnwise]{lineno}
\usepackage[compress,nospace]{cite}
\usepackage[dvipsnames]{xcolor}
\usepackage{CPL-2023}

\newcommand{\cplyear}{2026} \newcommand{\cplvol}{xx}
 \newcommand{\cplpagenumber}{xxxxxx}
 \newcommand{\cplpage}{\cplpagenumber-\thepage}

\begin{document}

\vspace*{-4mm}
\begin{center}
\large\bf{\boldmath{Rydberg-Atom Microwave Angle-of-Arrival Detection via Cylindrical Vapor-Cell-Mediated Field Redistribution}}
\footnotetext{\hspace*{-5.4mm}$^{*}$Corresponding authors. Email: renyuan\_821@aliyun.com; woohao@vip.163.com

$\dagger$These authors contributed equally to this work.}

\normalsize \rm{}Pei-Cheng Liu$^{1,2,\dagger}$, Xing-Chen Hu$^{1,3,\dagger}$, Yong Gao$^{1,4}$, Ao-Lin Guo$^{1,5}$, Yuan Ren$^{1,6,*}$, and Hao Wu$^{1,2,*}$
\\[3mm]\small\sl $^{1}$Lab of Quantum Detection \& Awareness, Space Engineering University, Beijing 101400, China

$^{2}$Department of Aerospace Science and Technology, Space Engineering University, Beijing 101400, China

$^{3}$School of Space Information, Space Engineering University, Beijing 101400, China

$^{4}$School of Space Command, Space Engineering University, Beijing 101400, China

$^{5}$Department of Basic Courses, Space Engineering University, Beijing 101400, China

$^{6}$National Key Laboratory of Space Target Awareness, Space Engineering University, Beijing 101400, China
\end{center}

\vskip 1.5mm

\small{\narrower Microwave angle-of-arrival (AoA) measurement is essential for radar, communication, and spectrum monitoring. Existing Rydberg-atom-based AoA schemes employ phase-difference measurements with local oscillators, standing-wave fluorescence imaging, or amplitude-ratio readout with internal metal reflectors. Here we demonstrate a new approach: a cylindrical glass vapor cell serving directly as an angle-encoding dielectric structure, eliminating the need for multiple apertures, local oscillators, or imaging optics. The cylindrical geometry produces angle-dependent reflection and field redistribution, mapping the incident AoA onto the effective microwave field sampled by the Rydberg ensemble. This effective field is read out optically via the Autler--Townes (A--T) splitting in the electromagnetically induced transparency (EIT) spectrum. Full-wave simulations and experiments at 11.64\,GHz confirm a deterministic, geometry-mediated response over $0^\circ$--$90^\circ$. Over the monotonic operating range ($55^\circ$--$125^\circ$), the angular resolution (minimum distinguishable increment) is $0.05^\circ$, and the angular accuracy (RMSE of repeated measurements) is $0.13^\circ$ across the full range, improving to $0.05^\circ$ in the optimal region ($65^\circ$--$115^\circ$). Occupying a sensing volume of $\sim$2.5\,cm$^3$, this method offers a compact, single-sensor pathway toward Rydberg AoA receivers.

\par}\vskip 3mm
{\it Introduction.} Atom-based electric field sensing has emerged as a powerful tool for radio-frequency, microwave, and terahertz metrology, benefiting from the intrinsic calibration and reproducibility of atomic transitions\ucite{1,2,3,4,5,6,7,8,9}. Electromagnetically induced transparency (EIT) involving highly excited Rydberg states enables all-optical readout of field-induced energy shifts and Autler--Townes (A--T) splitting, establishing a SI-traceable pathway for broadband electric field detection\ucite{3,4,5}. Following seminal demonstrations of Rydberg microwave electrometry, substantial progress in sensitivity enhancement\ucite{10,11,12,13}, portable sensor development\ucite{14,15}, and quantum-limited detection\ucite{16,17} has positioned Rydberg atoms as a compelling complement to conventional antenna technology.

Beyond scalar field metrology, Rydberg sensors are evolving into multifunctional quantum receivers capable of extracting amplitude, phase, polarization, and modulation information\ucite{18,19,20,21,22,23}. These developments have enabled applications in wireless communications, radar, and spectrum monitoring\ucite{24,25,26}. A particularly demanding task is angle-of-arrival (AoA) measurement, which requires the sensor to distinguish the direction of an incident wave with high precision. For this task, several distinct strategies have been explored. The first exploits the phase difference of arrival (PDOA) between two or more spatially separated quantum apertures within a single vapor cell, using heterodyne or superheterodyne readout with a local-oscillator (LO) reference field\ucite{27,28,29}. While effective, these methods require multi-channel photodetection, interferometric phase stability, and the emission of an LO field. The second replaces the LO with all-optical imaging of RF standing-wave fluorescence patterns, extracting the AoA from the spatial periodicity of the standing waves\ucite{30}; this all-passive solution demands a camera-based light-sheet imaging system and was demonstrated with an angular uncertainty of $\sim$1$^\circ$ and 11\,s integration time. A third approach embeds a metal plate inside a cylindrical vapor cell, generating a tailored standing-wave pattern whose amplitude ratio at two probe locations provides a monotonic AoA readout without LO or imaging optics\ucite{31}; the demonstrated angular resolution is approximately $1.7^\circ$. Computationally, magnitude-only power-profile methods employing RF-lens front ends have been proposed for multi-user AoA estimation\ucite{32}, though these remain at the simulation stage.

Among these approaches, the PDOA, standing-wave imaging, and PEC-plate methods each add an auxiliary element, such as a reference LO field, a camera imaging system, or an internal metal plate, to encode the AoA. Here we propose a strategy that dispenses with such auxiliary elements: we engineer the vapor cell geometry itself as the angle-encoding dielectric structure, so that the incident AoA is mapped onto the magnitude of the effective microwave field sampled by a single Rydberg ensemble. A cylindrical cell is chosen because its curved dielectric boundary produces AoA-dependent reflection, refraction, and internal field redistribution, generating a deterministic $\mathcal{E}_{\rm eff}$--$\theta$ response without requiring multiple apertures, LO fields, or imaging optics. By reading out this effective field via the A--T splitting of a single EIT probe, we establish a compact, all-optical AoA detection scheme. We validate this mechanism through combined full-wave HFSS simulations and experiments at a source--cell distance of 50\,cm with 0\,dBm incident power, and systematically evaluate the repeatability, error budget, and angular resolution of the method.

\medskip

{\it Experimental Setup and Principle.} Figure~1(a) shows the experimental setup and relevant energy levels. We employ a cesium ladder-type EIT system. An 852\,nm probe laser drives the $|6S_{1/2}\rangle \rightarrow |6P_{3/2}\rangle$ transition, and a 509\,nm coupling laser drives the $|6P_{3/2}\rangle \rightarrow |55D_{5/2}\rangle$ transition. The microwave field under test couples the Rydberg states $|55D_{5/2}\rangle \leftrightarrow |54D_{5/2}\rangle$. The probe and coupling wavelengths are 852.3473\,nm and 508.9538\,nm, respectively, and the microwave source is centered at 11.64\,GHz. Following two-photon excitation, atoms interact with the microwave field, producing A--T splitting in the probe transmission spectrum.

The cylindrical glass vapor cell (K9 glass, outer diameter 8\,mm, length 50\,mm, wall thickness 1\,mm) is positioned with the probe and coupling beams overlapping near its center. The probe beam has a power of 70\,$\mu$W and a $1/e^2$ radius of 0.8\,mm, and the coupling beam has a power of 30\,mW and a $1/e^2$ radius of 1.0\,mm. The microwave antenna is positioned 50\,cm from the cell center. At 11.64\,GHz, this distance satisfies the far-field condition for the horn antenna used\ucite{27}, ensuring that the incident wave front is approximately planar across the cell. The angle of arrival $\theta$ is defined as the angle between the antenna's line of sight and the cell center [Fig.~1(b)]. By rotating the antenna along a precision circular rail (radius 50\,cm, stepper-motor driven) while maintaining the same polarization, $\theta$ is varied from $0^\circ$ to $180^\circ$. Owing to the approximate rotational symmetry of the cylindrical cell about its transverse axis, the $\mathcal{E}_{\rm eff}$--$\theta$ response extends over the full $0^\circ$--$180^\circ$ range, and the sign of the calibration slope $d\mathcal{E}_{\rm eff}/d\theta$ distinguishes between the $0^\circ$--$90^\circ$ and $90^\circ$--$180^\circ$ intervals. The 852\,nm laser is frequency-stabilized via saturated absorption spectroscopy, delivered through an optical fiber, and combined with the modulated 509\,nm laser in the cell. The transmitted probe light is detected by a photodetector and acquired by a data acquisition card (100\,MHz sampling rate, 20,000 points per acquisition). A double-Lorentzian model is used to fit the A--T splitting. The setup is enclosed in an electromagnetic shielding box at $25\pm0.1\,^\circ$C to minimize thermal drift.

\vskip 4mm

\fl{1}\centerline{\resizebox{0.82\linewidth}{!}{\includegraphics{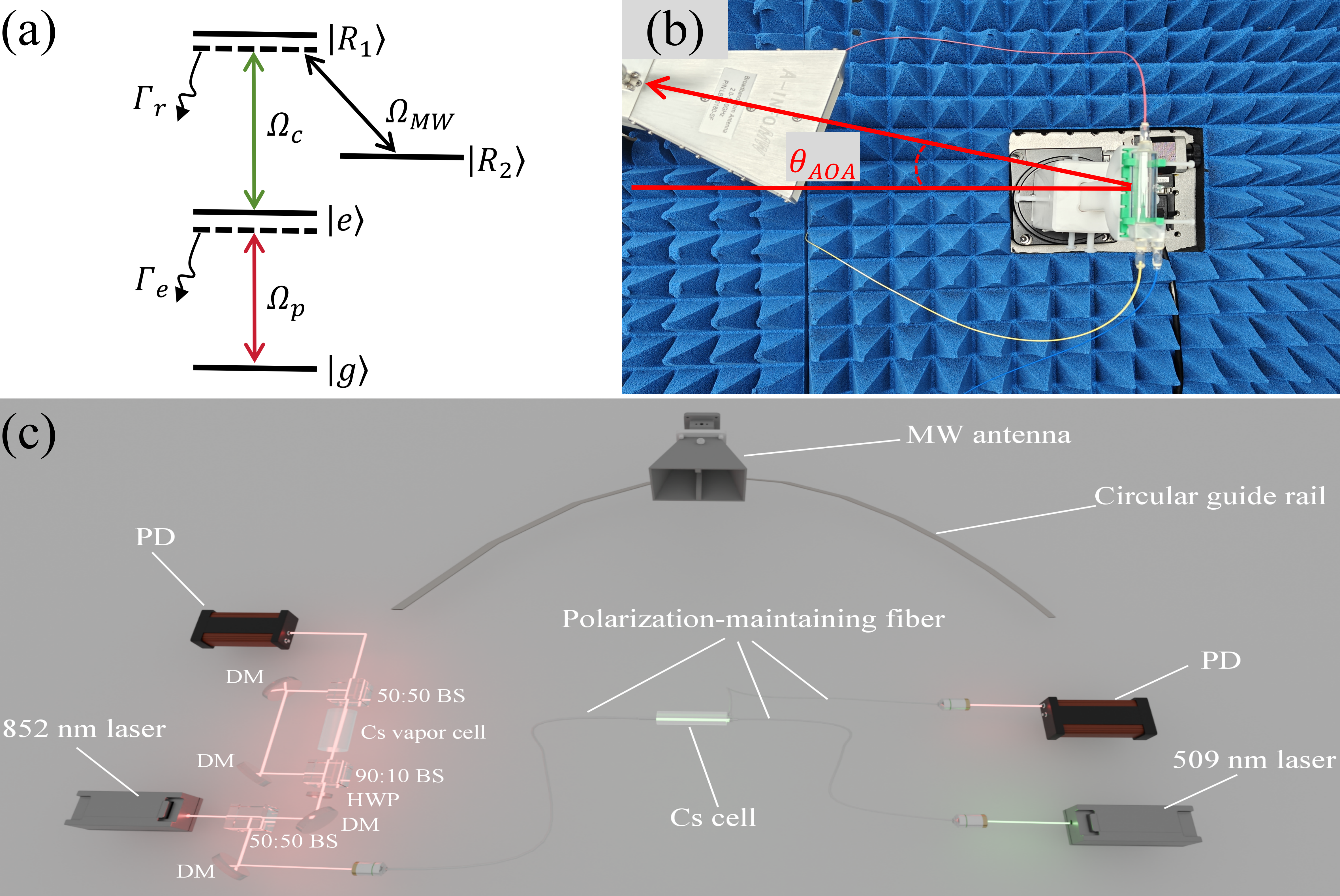}}}

\vskip 2mm

\figcaption{7.5}{1}{Experimental setup. (a) Energy-level diagram. (b) Definition of the microwave angle of arrival $\theta$. (c) Experimental configuration. DM: dichroic mirror; PD: photodetector; BS: beam splitter (T:R = 50:50 or 90:10); HWP: half-wave plate.}

\medskip

The microwave field enters the cell at an incident angle $\theta$. Owing to the dielectric boundary and curvature of the cylindrical cell, the incident field is partially reflected and redistributed, producing an angle-dependent microwave amplitude in the atomic sensing region. The fundamental relationship linking the spectroscopic readout to the local microwave field is given by the Rabi frequency of the Rydberg transition\ucite{10,30}:
\begin{eqnarray}\el{1}
\Omega_{\rm MW}(\mathbf{r}) = \frac{\mu_{\rm MW}\,|\mathbf{E}(\mathbf{r},\theta)|}{\hbar},
\end{eqnarray}
where $\mu_{\rm MW}$ is the transition dipole moment between the two Rydberg states and $|\mathbf{E}|$ is the field amplitude. The measured A--T splitting in frequency units is $\Delta f_{\rm AT} = \Omega_{\rm MW}/(2\pi)$. To account for the spatial averaging inherent to a thermal atomic ensemble, we define an effective field $\mathcal{E}_{\rm eff}$ over the sensing volume $V$:
\begin{eqnarray}\el{2}
\mathcal{E}_{\rm eff}(\theta) = \frac{\int_V |\mathbf{E}(\mathbf{r},\theta)|\, w(\mathbf{r})\, d^3\mathbf{r}}{\int_V w(\mathbf{r})\, d^3\mathbf{r}},
\end{eqnarray}
where $w(\mathbf{r})$ is the spatial weighting determined by the overlap of the probe and coupling beams. In the linear A--T regime, the measured splitting follows
\begin{eqnarray}\el{3}
\Delta f_{\rm AT}(\theta) = \frac{\mu_{\rm MW}}{h}\,\mathcal{E}_{\rm eff}(\theta).
\end{eqnarray}
Equation~(3) establishes the central encoding principle: the cylindrical cell maps $\theta$ onto $\mathcal{E}_{\rm eff}$, which is then read out through $\Delta f_{\rm AT}$. For each $\theta$, the probe transmission spectrum is recorded and fitted with a double-Lorentzian model to extract $\Delta f_{\rm AT}$. The effective field derived from $\Delta f_{\rm AT}$ is compared directly with the simulated value. RF-absorbing material surrounds the setup to suppress background reflections.

\medskip

{\it Results.} To elucidate the angle-dependent field redistribution, we first performed HFSS simulations replicating the experimental geometry. Figures~2(a)--2(d) show the simulated field distribution in the central cross-section of the cell at $\theta = 0^\circ$, $30^\circ$, $60^\circ$, and $90^\circ$. The internal field is strongly reshaped by the cylindrical dielectric boundary. At $\theta = 0^\circ$, the wave propagates nearly parallel to the cell axis, producing a predominantly forward-propagating field with weak scattering from the curved sidewall and a moderate peak near the rear wall due to the dielectric impedance mismatch. As $\theta$ increases to $30^\circ$, the wave begins to illuminate the sidewall at a grazing angle, generating asymmetric internal reflections that shift the field maxima toward the illuminated side. At $\theta = 60^\circ$, the incident wave illuminates both the sidewall and a portion of the end face, producing strong internal reflections that concentrate the microwave field near the boundary. At $\theta = 90^\circ$, the wave is incident on the cylindrical end face, producing a nearly symmetric standing-wave pattern along the axis, with the strongest field concentration at the center where the probe and coupling beams overlap. These four representative cases illustrate how the cylindrical geometry continuously redistributes the internal field magnitude in the atomic sensing volume as the incident angle changes. We define the effective aperture as the overlap volume of the probe and coupling beams where Rydberg excitation occurs [Fig.~2(e)]. The angle-dependent overlap between the internal field distribution and this finite sampling volume provides the physical basis for encoding the AoA onto the atomic response.

\vskip 4mm

\fl{2}\centerline{\resizebox{0.82\linewidth}{!}{\includegraphics{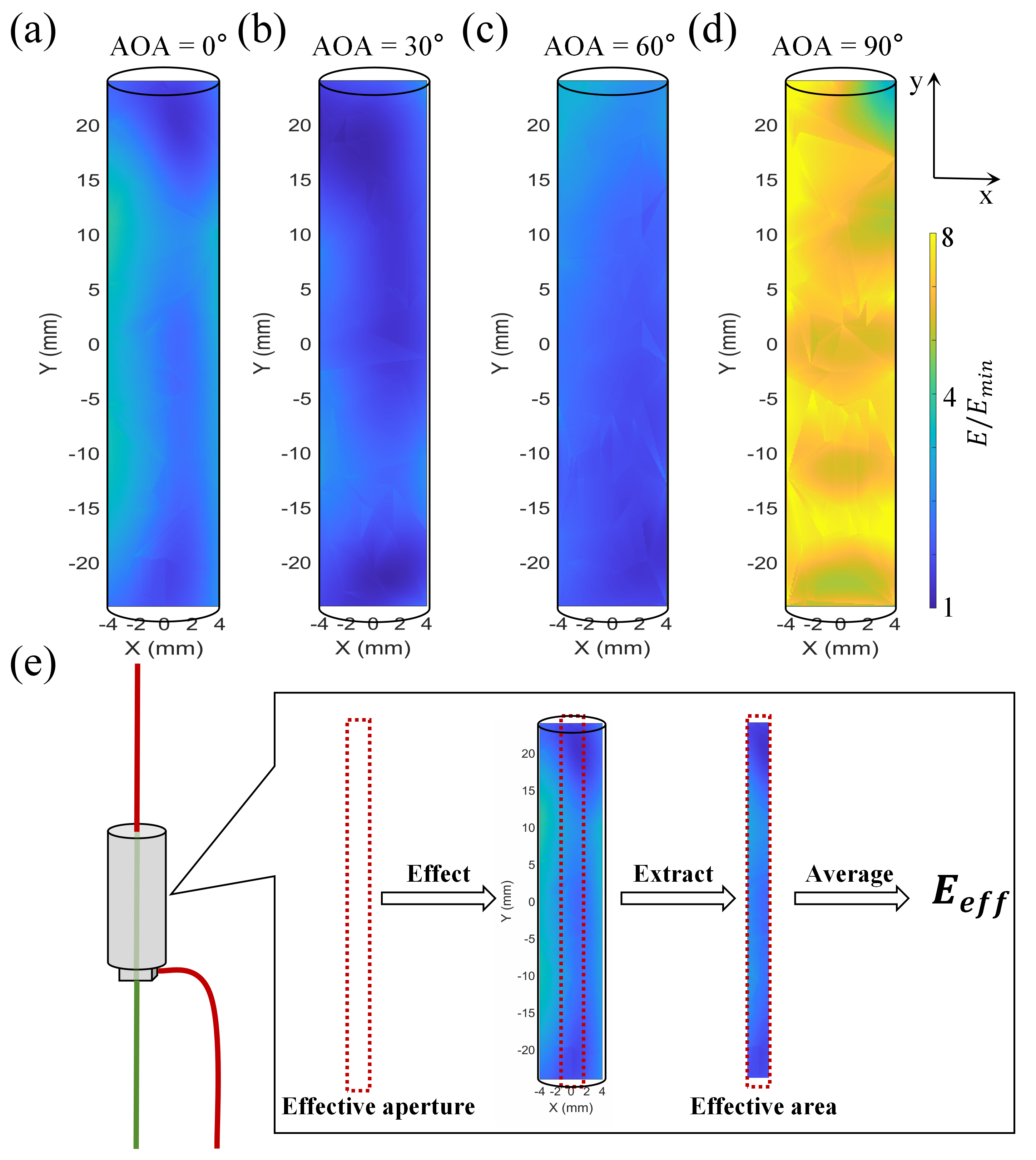}}}

\vskip 2mm

\figcaption{7.5}{2}{Simulated field distribution inside the cylindrical cell. (a)--(d) Maps at $\theta = 0^\circ$, $30^\circ$, $60^\circ$, $90^\circ$ ($D = 50$\,cm, $W = 0$\,dBm). (e) Left: photograph of the Cs cell; right: effective field calculation. The red dashed line indicates the effective aperture (probe--coupling overlap). All panels share the same color scale.}

\medskip

Figure~3(a) shows the normalized effective field $\mathcal{E}_{\rm eff}$ along the cell $y$-axis at $\theta = 0^\circ$, $30^\circ$, $60^\circ$, and $90^\circ$, obtained by weighted averaging over the $x$-direction. Several features of these profiles merit attention. First, at $\theta = 0^\circ$, the field along $y$ is dominated by the forward-transmitted wave with a monotonic decay away from the illuminated end. At $\theta = 30^\circ$, a pronounced peak appears near $y \approx 15$\,mm, corresponding to the constructive interference of the directly transmitted wave and the sidewall-reflected component. At $\theta = 60^\circ$, the profile becomes bimodal, with characteristic peaks arising from the superposition of the transmitted wave and internally reflected components off both the cylindrical sidewall and the end face. 
At $\theta = 90^\circ$, the wave is incident on the cylindrical end face, with the point of highest intensity concentrated at the end closest to the microwave source.
The peak-like features at fixed $\theta$ arise from the cylindrical geometry itself and from the beam-splitter mirror attached to one side of the cell, which introduces an additional reflective boundary.

Figure~3(b) presents $\mathcal{E}_{\rm eff}$ as a function of $\theta$ computed via Eq.~(2). With the cylindrical cell present (red curve), the response is piecewise: a gradual decrease over $0^\circ$--$20^\circ$, a rise over $20^\circ$--$35^\circ$, another decrease over $35^\circ$--$55^\circ$, and a monotonic increase over $55^\circ$--$90^\circ$ approaching unity. The piecewise structure originates from the interplay between the incident wave direction and the cylindrical geometry. At near-normal incidence ($\theta \approx 0^\circ$), the wave vector is nearly parallel to the cell axis, producing minimal reflection from the curved sidewall. As $\theta$ increases beyond $\sim$$20^\circ$, the projection of the wave vector onto the radial direction grows, leading to stronger internal reflections from the curved sidewall that modulate the local field amplitude. The transition near $55^\circ$ corresponds to the angle at which the dominant illumination shifts from the cylindrical sidewall to the end face, producing a qualitatively different internal field configuration and the monotonic response observed over $55^\circ$--$90^\circ$. In a control simulation with the cell removed (blue curve), the $\mathcal{E}_{\rm eff}$ variation is markedly reduced and dominated solely by the horn antenna's radiation pattern. The residual variation (within $\pm$0.05) in the control case is negligible compared to the $\sim$0.6 modulation produced by the cell, confirming that the angle-dependent response originates from cell-mediated field redistribution rather than geometric or antenna effects.

\vskip 4mm

\fl{3}\centerline{\resizebox{0.82\linewidth}{!}{\includegraphics{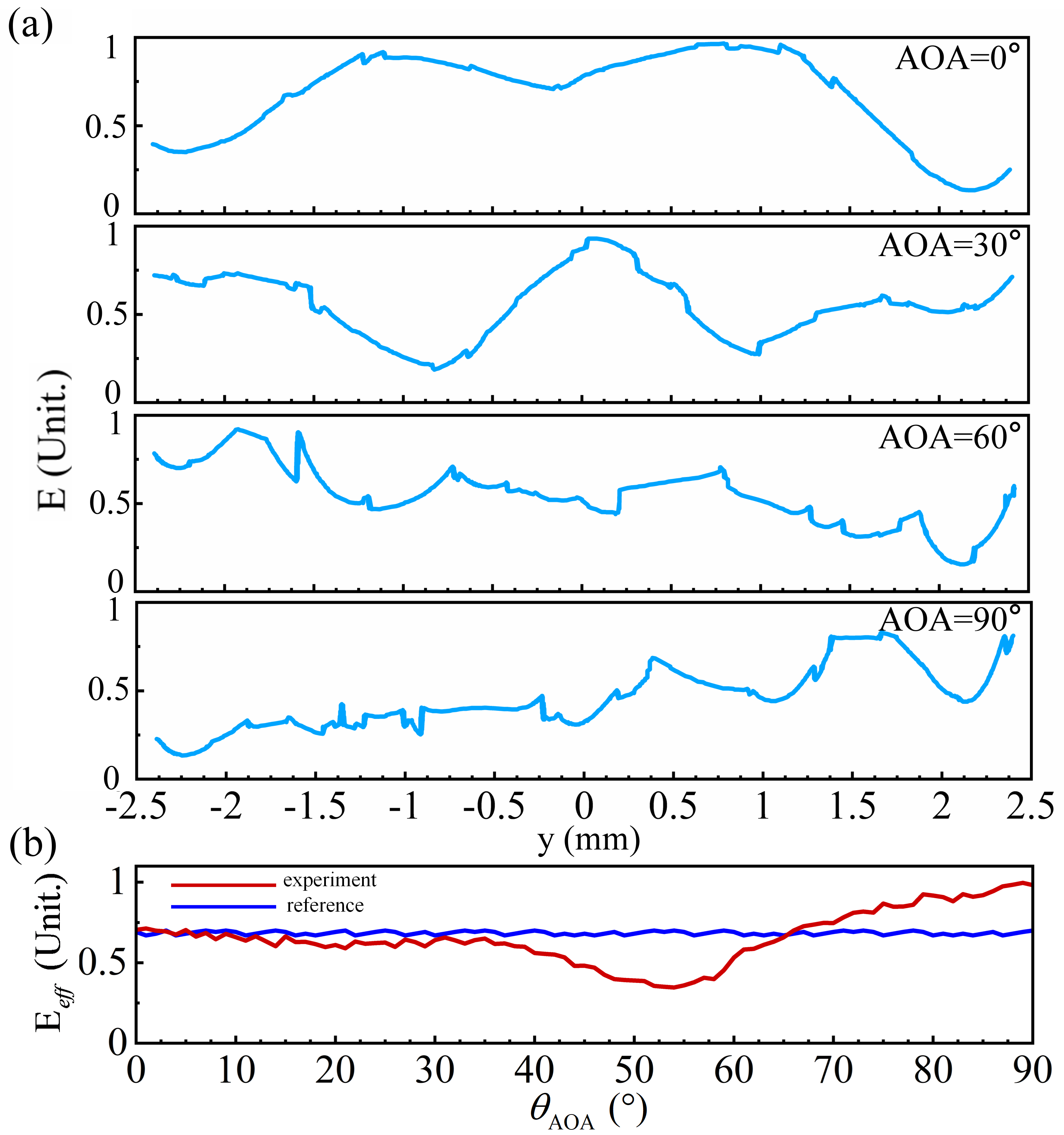}}}

\vskip 2mm

\figcaption{7.5}{3}{Normalized effective field $\mathcal{E}_{\rm eff}$ within the effective aperture. (a) Distribution along the cell $y$-axis at $\theta = 0^\circ$, $30^\circ$, $60^\circ$, $90^\circ$ ($x$-averaged). (b) $\mathcal{E}_{\rm eff}$ as a function of AoA. Red: with cell; blue: without cell (control).}

\medskip

Guided by the simulations, we measured the EIT spectra with microwave-induced A--T splitting at various incident angles. Figure~4(a) shows four representative probe transmission spectra at $\theta = 0^\circ$, $30^\circ$, $60^\circ$, and $90^\circ$ ($D = 50$\,cm, $W = 0$\,dBm). At $\theta = 0^\circ$, the A--T doublet is weakly separated, with the two transmission peaks partially overlapping. As $\theta$ increases to $30^\circ$, the splitting grows, and the two peaks become well resolved with a clear transmission dip between them. At $\theta = 60^\circ$, the splitting decreases. At $\theta = 90^\circ$, the splitting reaches its maximum, and the two peaks are fully separated, consistent with the maximum $\mathcal{E}_{\rm eff}$ observed in simulation. The consistent double-Lorentzian profile across all angles confirms that the system operates in the linear A--T regime, where the splitting is proportional to the field amplitude rather than being broadened by power effects.

Figure~4(b) compares the normalized $\mathcal{E}_{\rm eff}$ extracted from the measured $\Delta f_{\rm AT}$ via Eq.~(3) with the HFSS simulation. The experimental data track the simulated curve across the full angular range. A discrepancy near $70^\circ$--$80^\circ$ reflects the angular sensitivity of the sidewall reflection, where minor antenna alignment errors can shift the local field maximum. The overall agreement confirms that the spectral response is governed by the cell-mediated internal field redistribution described by Eq.~(3).

\vskip 4mm

\fl{4}\centerline{\resizebox{0.82\linewidth}{!}{\includegraphics{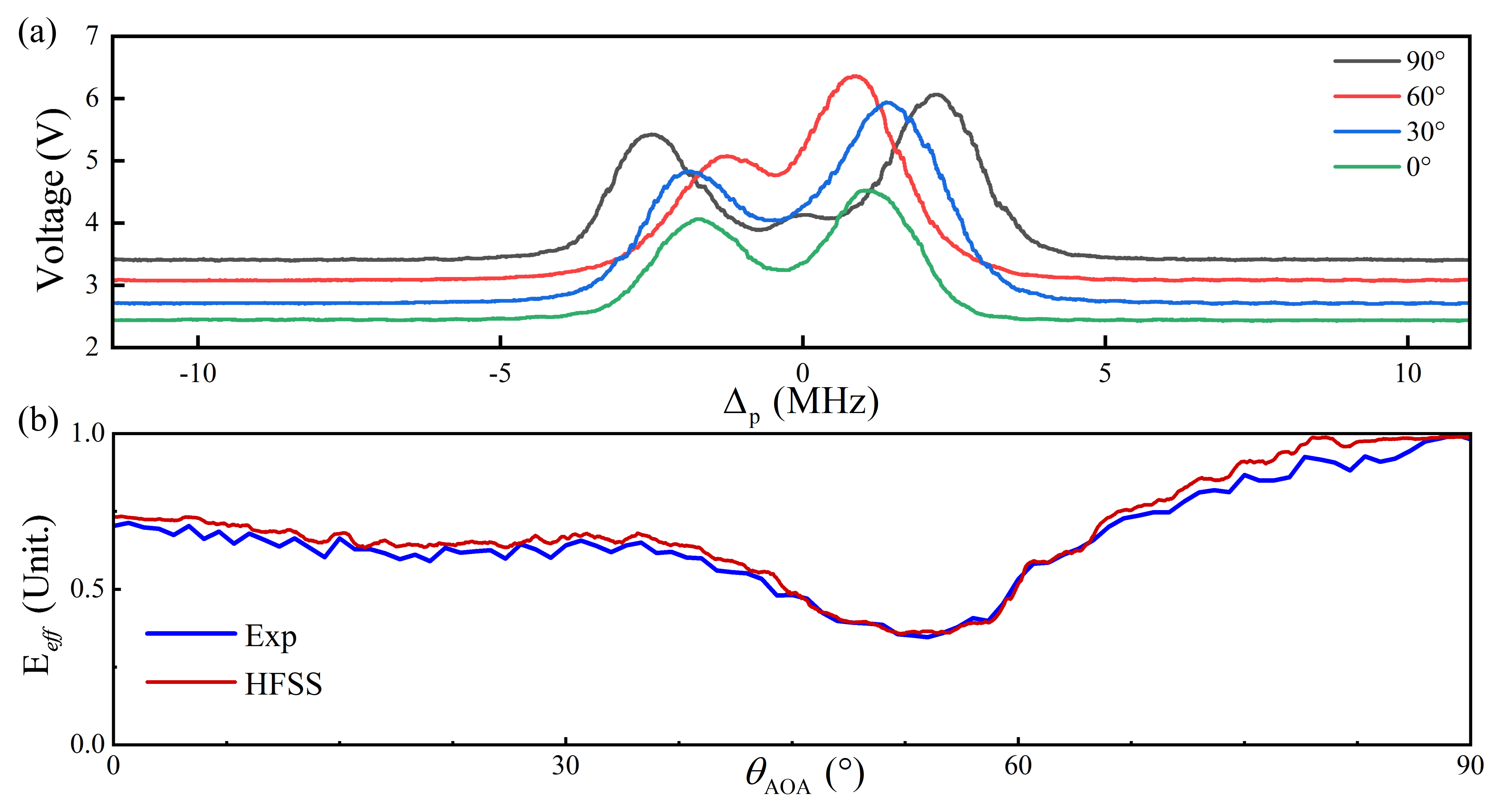}}}

\vskip 2mm

\figcaption{7.5}{4}{Experimental results. (a) Probe transmission spectra showing A--T splitting at $\theta = 0^\circ$, $30^\circ$, $60^\circ$, $90^\circ$ ($D = 50$\,cm, $W = 0$\,dBm; 20,000 points per spectrum, 100\,MHz sampling). (b) Normalized $\mathcal{E}_{\rm eff}$ as a function of AoA. Red symbols: $\mathcal{E}_{\rm eff}$ extracted from the measured $\Delta f_{\rm AT}$ via Eq.~(3); blue curve: HFSS simulation.}

\medskip

From Fig.~4(b), the $\mathcal{E}_{\rm eff}$--$\theta$ curve shows a weak dependence over $0^\circ$--$55^\circ$ (from $\sim$0.7 to $\sim$0.4, with a local rise near $30^\circ$) and a monotonic increase over $55^\circ$--$90^\circ$ (from $\sim$0.4 to $\sim$1.0). The weak dependence over $0^\circ$--$55^\circ$ arises because the cylindrical sidewall presents a nearly constant projected cross-section to the incident wave over this angular span. The monotonic segment over $55^\circ$--$90^\circ$ arises as the illuminated face transitions from the sidewall to the end face, where the projected cross-section changes rapidly with $\theta$. We therefore select the $55^\circ$--$90^\circ$ range as the primary operating range.

We then demonstrate AoA measurement over the extended range $55^\circ$--$125^\circ$ and evaluate the angular resolution. Twenty repeated measurements were performed by continuously rotating the antenna at $0.05^\circ$/s while recording the normalized field. Kernel ridge regression was employed to fit the $\mathcal{E}_{\rm eff}$--$\theta$ calibration curve. Figure~5(a) (upper panel) shows that $\mathcal{E}_{\rm eff}$ evolves smoothly over $55^\circ$--$125^\circ$, first increasing, reaching a maximum near $90^\circ$, and then decreasing with a response curve that is approximately symmetric about $90^\circ$ due to the rotational symmetry of the cylindrical cell. Consequently, two distinct incident angles may produce the same $\mathcal{E}_{\rm eff}$ value. The sign of $d\mathcal{E}_{\rm eff}/d\theta$ resolves this ambiguity, being positive for $55^\circ < \theta < 90^\circ$ and negative for $90^\circ < \theta < 125^\circ$.  Figure~5(b) displays the standard deviation of $\mathcal{E}_{\rm eff}$: 0.0013--0.0026 over $55^\circ$--$65^\circ$ and $115^\circ$--$125^\circ$, and below 0.0013 over $65^\circ$--$115^\circ$, with the best performance near $90^\circ$. The higher uncertainty at the edges arises from the reduced slope of the calibration curve near the turning points.

\vskip 4mm

\fl{5}\centerline{\resizebox{0.82\linewidth}{!}{\includegraphics{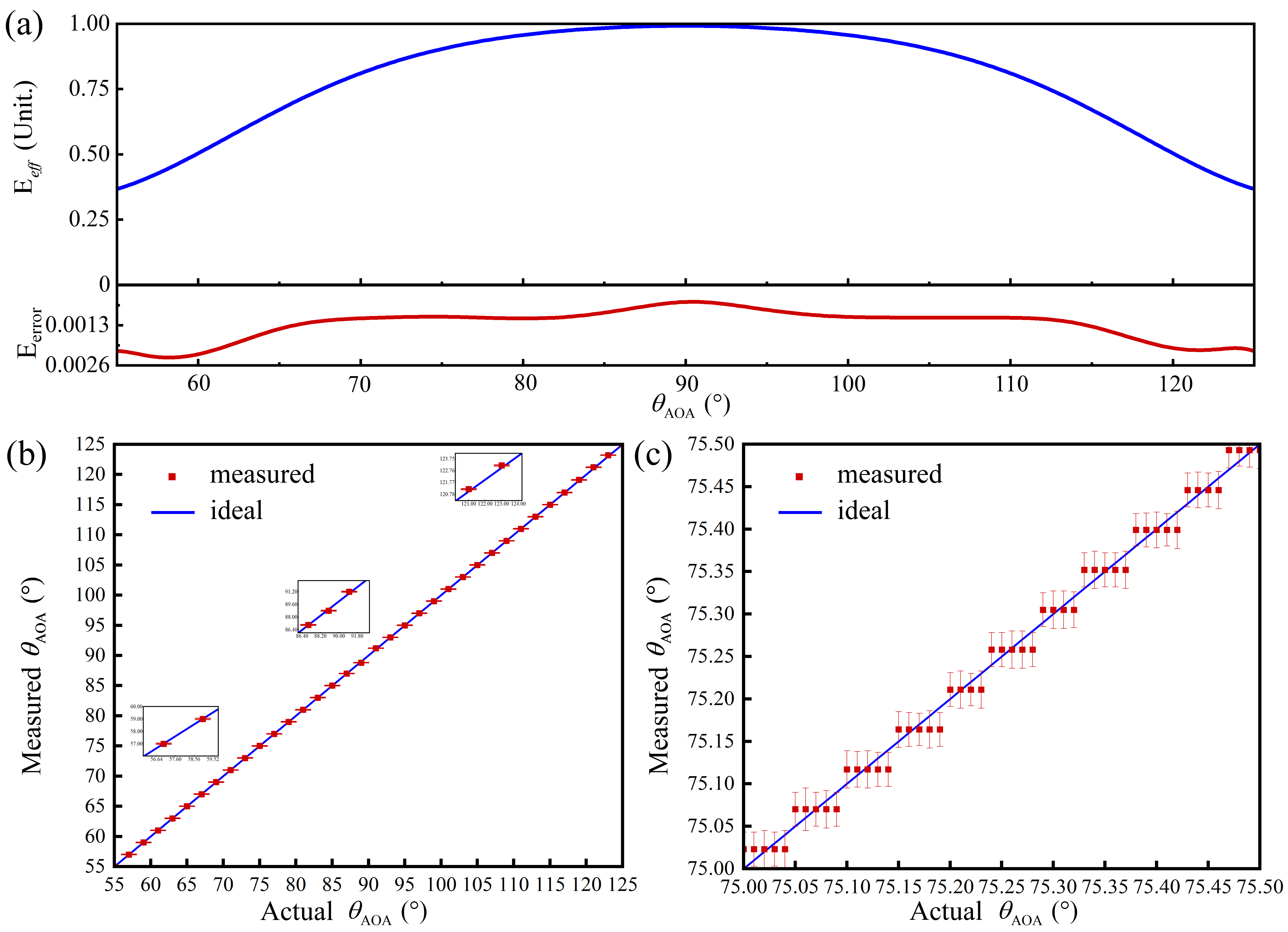}}}

\vskip 2mm

\figcaption{7.5}{5}{Repeatability measurements. (a) Upper: effective field as a function of AoA (kernel ridge regression fit); lower: standard deviation as a function of AoA (20 repeated scans). (b) Measured AoA against true AoA ($55^\circ$--$125^\circ$). Red markers: experiment; blue line: ideal. Full-range RMSE: $0.13^\circ$; optimal region ($65^\circ$--$115^\circ$) RMSE: $0.05^\circ$. (c) High-resolution view over $75^\circ$--$75.5^\circ$, showing $0.05^\circ$ angular resolution (ten discrete levels).}

\medskip

Figure~5(b) compares the measured and true AoA values over $55^\circ$--$125^\circ$. The root-mean-square deviation (RMSE) between the measured and true angles is $0.13^\circ$ over the full range. The performance is notably better in the central region ($65^\circ$--$115^\circ$), where the RMSE improves to $0.05^\circ$, matching the DAQ quantization limit. The larger deviations at the angular edges ($55^\circ$--$60^\circ$ and $120^\circ$--$125^\circ$) arise from the reduced slope of the calibration curve, which amplifies the effect of field-readout noise as described by $\delta\theta \propto |d\mathcal{E}_{\rm eff}/d\theta|^{-1}$. Figure~5(c) shows a detailed view over $75^\circ$--$75.5^\circ$, where the data segregate into discrete horizontal stripes spaced by $0.05^\circ$. Between $75.0^\circ$ and $75.5^\circ$, ten such levels are resolved, demonstrating the $0.05^\circ$ angular resolution of the combined acquisition-and-fitting pipeline.

\medskip

{\it Discussion.} The angular resolution and accuracy are determined by distinct factors. The resolution, defined as the minimum distinguishable angular increment, is $0.05^\circ$ and follows from the discrete step structure observed in Fig.~5(c). The accuracy, defined as the deviation between measured and true AoA values, is characterized by the RMSE of $0.13^\circ$ over the full $55^\circ$--$125^\circ$ range, improving to $0.05^\circ$ in the optimal region ($65^\circ$--$115^\circ$). The full-range RMSE exceeds the optimum because the calibration slope is shallower at the angular edges ($55^\circ$--$65^\circ$ and $115^\circ$--$125^\circ$), where the same level of field-readout noise translates to a larger angular uncertainty. In the optimal region, the accuracy equals the DAQ resolution floor, indicating that the spectroscopic noise lies at or below the current hardware limit and that further improvements in angular accuracy would require upgrading the digitization.

The architecture of the present scheme differs from previously reported Rydberg AoA methods. Phase-difference approaches\ucite{27,29} and array-based localization\ucite{28} rely on multi-channel readout with a local-oscillator reference; standing-wave fluorescence imaging\ucite{30} employs camera-based detection in a rectangular cell; the PEC-plate approach\ucite{31} uses an internal metal reflector and the amplitude ratio between two probe locations within a cylindrical cell; and the RF-lens method\ucite{32} remains a computational proposal. Our approach explores a complementary direction: a single-probe, auxiliary-element-free architecture in which the AoA is encoded into the scalar field amplitude solely through the engineered cell geometry. This architecture trades phase-based encoding for amplitude-based encoding; the resulting sensitivity to amplitude noise in the laser intensity and detection chain may make this approach best suited to platforms where phase-noise limitations currently dominate.

Several aspects of the present demonstration suggest directions for future improvement. On the hardware side, the $0.05^\circ$ angular resolution and optimal-region accuracy are both DAQ-limited; a higher-resolution digitizer would enable the physical noise floor to be assessed. On the geometry side, the encoding sensitivity can be enhanced through cell optimization. The present cell provides sufficient optical depth for the EIT readout while maintaining a pronounced angular response from the curved sidewall. A larger diameter-to-wavelength ratio would increase the contrast between the field strengths at different incident angles, steepening the $\mathcal{E}_{\rm eff}$--$\theta$ calibration curve and thereby improving the achievable angular accuracy. Reducing the length-to-diameter ratio or introducing an asymmetric end cap could further sharpen the angular dependence by modifying the internal reflection geometry. On the architecture side, the $180^\circ$ ambiguity inherent to a symmetric cylindrical cell could be resolved by a symmetry-breaking element, such as an off-axis probe-coupling arrangement or a partial dielectric coating. Extending the method to two-dimensional AoA (elevation and azimuth) could be achieved with a multi-region interrogation scheme in which two orthogonally configured probe pairs sample independent angular encoding functions.

\medskip

{\it Conclusion.} We have demonstrated microwave AoA detection using a single Rydberg-atom sensor, exploiting angle-dependent field redistribution within a cylindrical glass vapor cell. Full-wave simulations reveal that the dielectric boundary reshapes the internal microwave field as a function of incident angle. The A--T splitting readout establishes a deterministic relationship between the AoA and the measured spectrum. Experiments at a source--cell distance of 50\,cm with 0\,dBm incident power validate the angle-encoding mechanism. The angular resolution reaches $0.05^\circ$, and the angular accuracy (RMSE) is $0.13^\circ$ over the full $55^\circ$--$125^\circ$ range, improving to $0.05^\circ$ in the optimal region where the calibration slope is steepest. In the optimal region, the accuracy equals the DAQ resolution floor, indicating that the spectroscopic noise floor lies at or below the current hardware limit. This method provides a single-sensor, all-optical pathway for atomic AoA detection within a sensing volume of $\sim$2.5\,cm$^3$, distinct from both multi-receiver phase-difference and standing-wave fluorescence imaging approaches. Further improvements are anticipated through cell geometry optimization and higher-fidelity readout.

\medskip

\textit{Acknowledgements.} This work was supported by the National Natural Science Foundation of China (62173342, 62203466, 12301533, 62302082).

\end{document}